\documentclass[11pt]{article}
\usepackage{amsmath,amssymb,color,graphics,epsfig}
\usepackage{amsfonts}
\newcommand{\be}{\begin{equation}}
\newcommand{\ee}{\end{equation}}
\newcommand{\bea}{\setlength\arraycolsep{2pt} \begin{eqnarray}}
\newcommand{\eea}{\end{eqnarray}}
\newcommand{\nn}{\nonumber}
\newcommand{\mc}{\mathcal}
\newcommand{\mm}{\mathrm}
\newcommand{\pa}{\partial}

\def\ft#1#2{{\textstyle{\frac{\scriptstyle #1}{\scriptstyle #2} } }}
\def\fft#1#2{{\frac{#1}{#2}}}

\def\0{{\sst{(0)}}}
\def\1{{\sst{(1)}}}
\def\2{{\sst{(2)}}}
\def\3{{\sst{(3)}}}
\def\4{{\sst{(4)}}}
\def\5{{\sst{(5)}}}
\def\6{{\sst{(6)}}}
\def\7{{\sst{(7)}}}
\def\8{{\sst{(8)}}}
\def\sst#1{{\scriptscriptstyle #1}}

\begin{document}

\begin{flushright}
%\hfill{KIAS-P12028}
 %\hfill{
%\bf hep-th/yymmnnn}
\end{flushright}

\vspace{25pt}
\begin{center}
{\large {\bf Criticality of central charges for Gauss-Bonnet black holes}}

\vspace{10pt}
 Hong-Ming Cui$^{1\dagger}$ and Zhong-Ying Fan$^{1\dagger}$

\vspace{10pt}
$^{1\dagger}${ Department of Astrophysics, School of Physics and Material Science, \\
 Guangzhou University, Guangzhou 510006, P.R. China }\\
%\smallskip
%{\it $^{2}$Department of Physics and State Key Laboratory of Nuclear Physics and Technology,\\}
%{\it Peking University, No.5 Yiheyuan Rd, Beijing 100871, P.R. China\\}
%\smallskip
%{\it $^{3}$Collaborative Innovation Center of Quantum Matter, No.5 Yiheyuan Rd,\\}
%{\it  Beijing 100871, P. R. China\\}

\vspace{40pt}

\underline{ABSTRACT}
\end{center}
Employing extended phase space formalism, we study critical phenomenon of A-charge and C-charge for holographic theories dual to Gauss-Bonnet black holes. We find an universal critical Gauss-Bonnet coupling, giving rise to an universal ratio between the two central charges at the critical point. This leads to a new intepretation for critical behavior of Gauss-Bonnet black holes in terms of the boundary degrees of freedoms, although the solutions are electrically neutral. Another novel feature is for either of the central charges, the transition temperature is beyond the critical point but is upper bounded by causality of the boundary theories.

%\vfill {\footnotesize  Email: fanzhy@gzhu.edu.cn\,.}

\thispagestyle{empty}

\pagebreak

\tableofcontents
\addtocontents{toc}{\protect\setcounter{tocdepth}{2}}

%%%%%%%%%%%%%%%%%%%%%%%%%%%%%%%%%%%%%%%%

%\newpage
%%%%%%%%%%%%%%%%%%%%%%%%%%%%%%%%%%%%%%%%

%\vspace{2cm}

\section{Introduction}
Since discovery of four laws of black hole thermodynamics \cite{Bekenstein:1972tm,Bekenstein:1973ur,Bardeen:1973gs}, numerous studies have been conducted for thermodynamic phase transition for a variety of black hole solutions. Of these works, black holes asymptotic to anti-de sitter spacetime (AdS) are of paramount importance according to AdS/CFT correspondence \cite{Maldacena:1997re,Witten:1998qj,Gubser:1998bc}. In \cite{Kubiznak:2012wp}
, the first law of charged AdS black holes was first extended to include thermodynamic pressure $P$ and volume $V$ by taking the cosmological constant $\Lambda$ as a thermodynamic variable. It was establised that the charged AdS black hole exhibits P-V criticality in the extended phase space, akin to the critical phenomenon of Van-der Waals fluids. The remarkable results facilitated numerous research for P-V criticality of a number of AdS black holes, typlically including rotating black holes \cite{Gunasekaran:2012dq,Cheng:2016bpx} and the Gauss-Bonnet black holes \cite{Cai:2013qga} that we are interested in in this work. 

Recently, the thermodynamics of charged AdS black holes was reexamined \cite{Cong:2021fnf,Cong:2021jgb} in the context of gauge-gravity duality. In this new setting, both $\Lambda$ and the Newton's gravitational constant $G$ were taken as variables in the bulk, resulting to a new first law extended with an extra pair of thermodynamic conjugate: the central charge $C$ in the boundary CFT and the dual chemical potential $\mu$. It was revealed \cite{Cong:2021fnf,Cong:2021jgb} that the Van der Waals-like behaviors of charged AdS black holes can be incorporated into this new pair of conjugates. Accordingly, this gives the first example to interpret critical phenomenon of AdS black holes in terms of the degrees of freedoms (dofs) in the boundary theory: the ratio of the number of neutral dofs $N\propto\sqrt{C}$ to that of charged carriers $Q$ plays a pivotal role to guarantee emergence of the Van-der Waals like behavior.

%However, this new understanding is virtually obviated in the originial setting of extended phase space [xxx].

In \cite{Kumar:2022afq}, the central charge criticality for Gauss-Bonnet black holes was investigated. However, the central charge introduced in that work is not dual to either of the central charges in the boundary CFTs. In holography, Gauss-Bonnet black holes in five dimensions are dual to four dimensional CFTs which generally have two distinct central charges, referred to as A-charge and C-charge. Inspired by this, we would like to work in a different context by varying both the Newton's gravitational constant and the Gauss-Bonnet coupling. The bulk first law is extended to include the variations of both the A-charge and the C-charge. As a consequence, we find an universal critical Gauss-Bonnet coupling, which corresponds to an universal ratio bewteen the two central charges at the critical point. This leads us able to inteprete the Van-der Waals like behaviors of Gauss-Bonnet black holes in terms of the boundary dofs, similar to the charged AdS black holes case.

The paper is orgainized as follows. In section 2, we brieflly review the thermodynamics of Gauss-Bonnet black holes and extend it with both A-charge and C-charge as well as their conjugates. In section 3, we study criticality of the central charges in detail. We derive a universal critical Gauss-Bonnet coupling, equivalent to a universal ratio between the two central charges. We also demonstrate that for both charges the transition temperature is beyond the critical point but is upper bounded by causality of the boundary theories. In section 4, we derive various critical exponents for central charge criticality. We conclude in section 5.

\section{Extended thermodynamics of Gauss-Bonnet black holes}\label{sec2}

Consider Gauss-Bonnet gravity in five dimension, whose Lagrangian density is given by
\be \mathcal{L}=\fft{1}{16\pi G}\Big[R+12\ell^{-2}+\fft{1}{2}\lambda\ell^2 \big(R^2-4R^2_{\mu\nu}+R^2_{\mu\nu\rho\sigma} \big)\Big] \,,\ee
where $\ell$ stands for the bare AdS radius and $\lambda$ is the dimensionless Gauss-Bonnet coupling. Notably, causality of the dual field theories constrains  $0<\lambda<9/100$ \cite{Brigante:2007nu,Camanho:2009vw}. This is the focal region throughout this work. The spherically symmetric black hole solution reads
\bea
&&ds^2=-f(r)dt^2+dr^2/f(r)+r^2 d\Omega^2_3\,, \nn\\
&&f(r)=1+\fft{r^2}{2\lambda\ell^2}\Big[1-\sqrt{1-4\lambda+\ft{32\lambda\ell^2 G M}{3\pi r^4}}\, \Big] \,,
\eea
where $d\Omega^2_3$ denotes the metric of a unit $3$-sphere and $M$ stands for the black hole mass. The event horizon is defined by the largest real root of the equation $f(r_h)=0$. Using standard method, the mass, the entropy and the temperature can be evaluated as
%\be T=\fft{\kappa}{2\pi}\,,\qquad S=-\ft 18 \int_+  d^{D-2}x\,\sqrt{h}\, \epsilon^{\mu\nu}\epsilon^{\rho\sigma}\, \fft{\partial L}{\partial R^{\mu\nu\rho\sigma}} \,,\ee
\bea
&&M=\fft{3\pi( r_h^4+r_h^2\ell^2+\lambda\ell^4 )}{8G\ell^2}\,,\\
&&S=\fft{\pi^2 r_h^3}{2G}\Big(1+\fft{6\lambda\ell^2 }{r_h^2} \Big) \,,\\
&&T=\fft{r_h(2r_h^2+\ell^2) }{2\pi \ell^2 (r_h^2+2\lambda\ell^2 )}\,.\label{tem}
\eea
It is straightforward to test that the first law of thermodynamics $dM=T dS$ holds. However, we are interested in the so-called {\it extended phase space}, in which new pairs of thermodynamic conjugates are encompassed. Perhaps the mostly known example is the thermodynamic pressure $P$
\be P=\fft{6\ell^{-2}}{8\pi G}\,,\label{pressure}\ee
and the thermodynamic volume $V$ since the pioneer work \cite{Kubiznak:2012wp}. Numerous studies have been conducted for critical phenomenon associated to this pair of variables, referred to as P-V criticality, for a variety of AdS black holes. However, in that case, the Newton's gravitational constant is usually held fixed so that central charge dual to the boundary CFT is disregarded. 

In this work, we will study the extended thermodynamics with two central charges, dubbed by A-charge and C-charge, for holographic theories dual to Gauss-Bonnet gravity. The two central charges can be evaluated via holograpic machine as \cite{Nojiri:1999mh,Myers:2010jv}
\bea
&& A=\fft{\pi^2\ell_{\mm{eff}}^3}{8\pi G}\big( 1-6\lambda f_\infty\big)\,,\label{acharge}\\
&& C=\fft{\pi^2\ell_{\mm{eff}}^3}{8\pi G}\big( 1-2\lambda f_\infty\big)\,,\label{ccharge}
\eea
where $\ell_{\mm{eff}}$ stands for the effective AdS radius, given by
\be f_\infty=\fft{\ell^2}{\ell^2_{\mm{eff}} }=\fft{1-\sqrt{1-4\lambda}}{2\lambda} \,.\ee
Notice that the central charges are dimensionless as well as the black hole entropy (we work in the units $\hbar=c=k_B=1$). With these new variables, the extended first law reads
\be dM=TdS+V dP+\mu_A dA+\mu_C dC \,,\ee
where $V$ is a novel thermodynamic volume
\be V=\fft{\pi^2}{10}\big( r_h^4+r_h^2\ell^2+\lambda \ell^4 \big) \,,\ee
and ($\mu_A$, $\mu_C$) recode the chemical potentials corresponding to the A-charge and C-charge respectively (these quantities have lengthy expressions, which are not intructive to present). The extended bulk first law matches with the boundary first law altered by the two central charges \cite{Dutta:2022wbh,Punia:2023ilo}. In addition, the standard scaling dimensional argument gives rise to the Smarr relation
\be M=\fft32\,(T S+\mu_A A+\mu_C C)-P V\,.\ee

To study whether the dual theory exhibits central charge criticality, we shall first rewrite the ``constants'' $(\ell\,,G\,,\lambda)$ in terms of thermoynamic variables $(P\,,C\,,A)$
\bea
&& \ell=\fft{2^{3/10}3^{1/5}\sqrt{3C-A}}{\pi^{2/5}P^{1/5}(5C-A)^{3/10}} \,,\label{bareradius}\\
&& G=\fft{3^{3/5}(5C-A)^{3/5}}{4\pi^{1/5}(2P)^{3/5}(3C-A)}\,, \label{newton}\\
&& \lambda=\fft{(5C-A)(C-A)}{4(3C-A)^2} \,,\label{lamratio}
\eea
where we have picked up the physical branch solutions corresponding to $\lambda>0$ or equivalently $C>A$. In fact, this implies a upper limit for the transition temperature, as will be detailed below. By plugging above results into (\ref{tem}), we obtain the equation of state $T=T(\mu_A\,,\mu_C\,,A\,,C)$, which enables us to explore critical phenomenon of the central charges. Before moving to details, we shall point out that P-V criticality disappears for the novel thermodynamic volume in canonical ensemble (fixing both central charges). Accordingly, the P-V criticality for Gauss-Bonnet black holes established previously \cite{Cai:2013qga} can be viewed as remanet of central charges criticality, which is incorporated into the original thermodynamic volume by taking proper scaling limit. %$A P^{3/2}\rightarrow \mm{cons}$.%

\section{Central charges criticality}

To study central charges criticality, we work in either the $(P\,,C)$ ensemble or the $(P\,,A)$ ensemble. Notably, in both cases, evaluating the derivatives $T'(r_h)\,,T''(r_h)$ yields an universal critical Gauss-Bonnet coupling
\be \lambda_c=\fft{1}{36} \,.\label{critlambda}\ee
This implies an universal ratio for the two central charges at the critical point per the relation (\ref{lamratio}). Intriguingly, it is very similar to the charged AdS black holes case \cite{Cong:2021fnf,Cong:2021jgb}, in which the ratio of the central charge to the square of electric charges is universal at the critical piont. Hence in both cases, the Van-der Waals like behaviors of black holes can be interpreted in terms of the boundary dofs. By comparison, the critical Newton's constant depends on the pressure and is non-universal, as can be seen in (\ref{newton}). This gives rise to a family of critical points as $G$ can be varied independently. The critical temperature and horizon radii are given by
\be T_c=\fft{\sqrt{6}}{2\pi\ell_c}\,,\quad r_{h\,,c}=\fft{\ell_c}{\sqrt{6}} \,,\ee
where $\ell_c$ denotes the critical bare AdS radius, which depends on the ensemble we opt for.
\begin{figure}
  \centering
  \includegraphics[width=210pt]{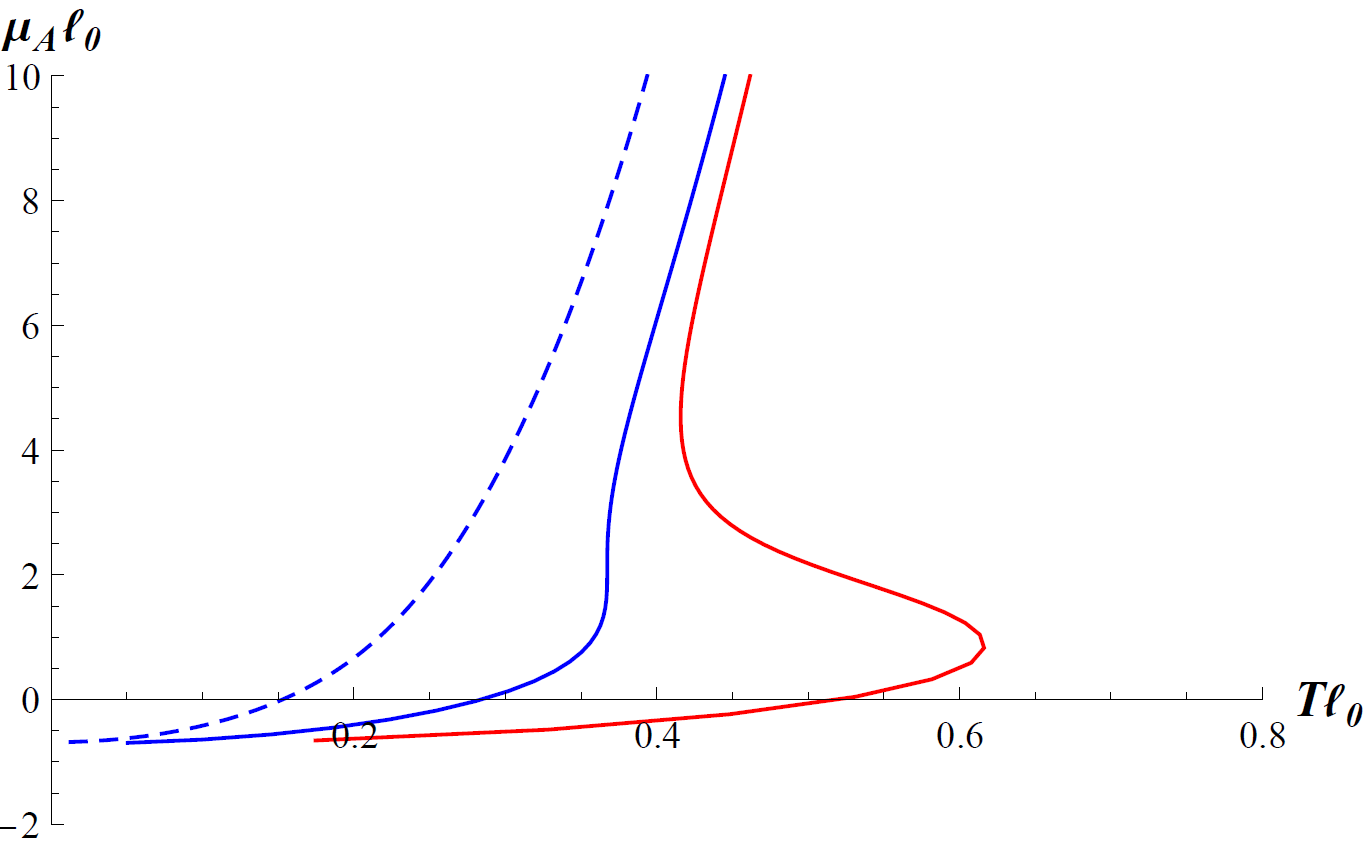}
  \includegraphics[width=210pt]{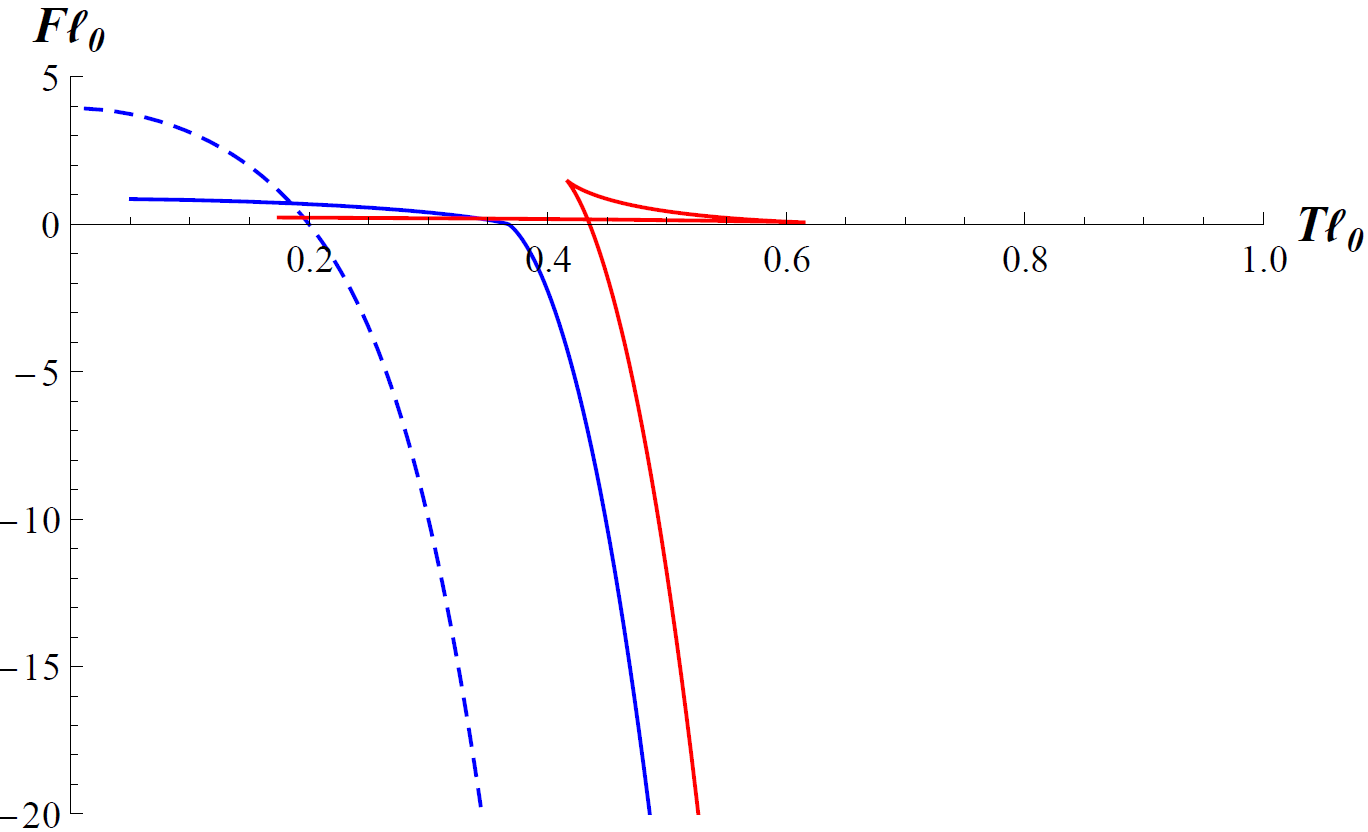}
  \caption{A-charge criticality. Left panel: $\mu_A-T$ diagram. $A/A_c=11/10$ (red), $A/A_c=1$ (blue) and $A/A_c=1/2$ (dashed). For $A>A_c$, the chemical potential $\mu_A$ has multi-values for a certain regime of temperatures. Right panel: $F-T$ diagram for the same parameters. Characteristic feature of first order transitions: the swallow tail behavior, emerges when $A>A_c$. We have set $P\ell_0^5=5\,,C=10$, where $\ell_0$ is an arbitrary reference length scale.}
  \label{acrit}
\end{figure}

\subsection{A-charge criticality}
According to holographic c-theorem \cite{Myers:2010tj}, the A-charge in higher dimensions can be regarded as a direct generalisation of the central charge in two dimensional CFTs. This inspires us to discern $\mu_A-A$ criticality at first. We opt to work in the $(P\,,C)$ ensemble, corresponding to fixing volume and the $C$-charge in the boundary conformal field theories. 

In addition to the critical temperature and horizon radii, we obtain a critical A-charge and bare AdS radius
\bea
&& A_c=\fft{3(2-\sqrt{2})}{2}\,C\,,\nn\\
&& \ell_c=\fft{2^{7/20} 3^{7/10 } C^{1/5} }{ \pi^{2/5}(4+3\sqrt{2})^{3/10} P^{1/5} }\,.
\eea
Notice that by varying the pressure $P$, we can obtain a family of critical point. However, existence of a critical point in the parameters space neither necessarily implies critical phenomenon accordingly nor any detail about it. To unravel this, we show $\mu_A-T$ diagram for various A-charges in the left panel of Fig. \ref{acrit}. It is immediately seen that for $A>A_c$, the chemical potential $\mu_A$ has multi-values for a certain regime of temperatures. This implies the presence of the first order transition. Besides, by showing the free energy $F=M-T S$ as a function of temperature, we reveal swallower tail behavior for $A>A_c$, as shown in the right panel of Fig. \ref{acrit}. These features enable us to confirm first order transition occurs between a pair of black holes for the central charge beyond the critical point.

\begin{figure}
  \centering
  \includegraphics[width=250pt]{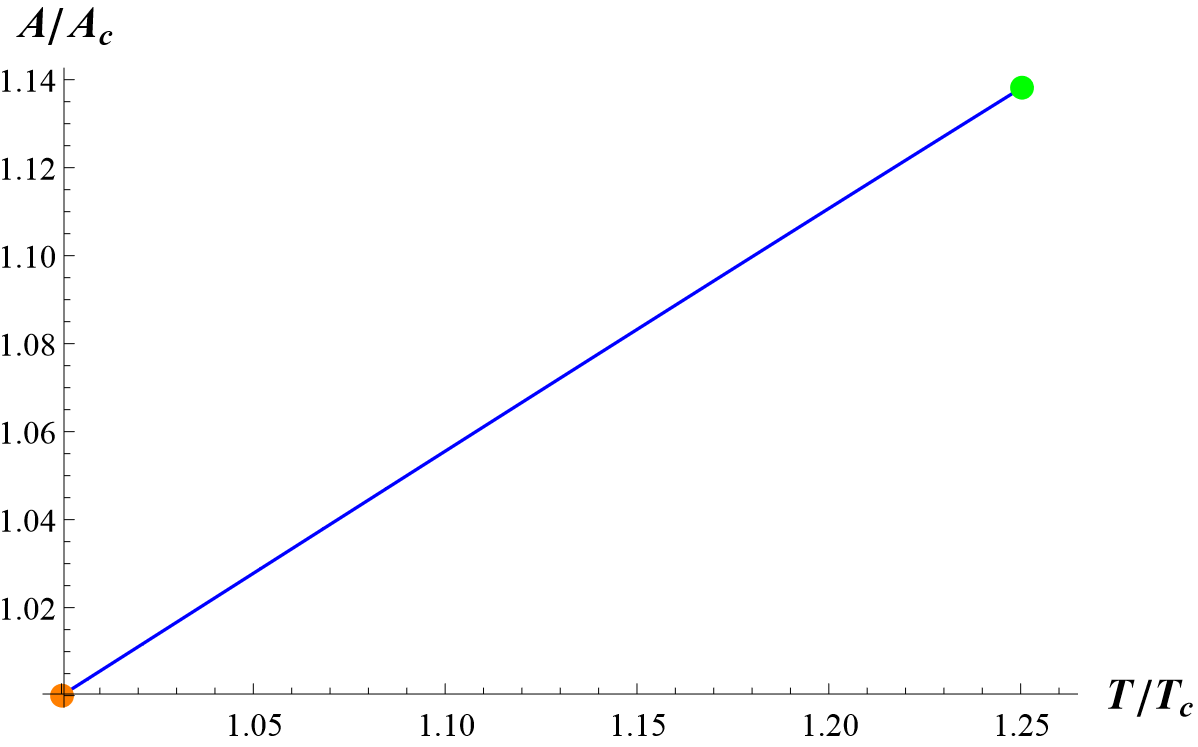}
  \caption{Coexistence curve for A-charge criticality. The solid circles stand for the critical point (orange) and the maximum transition temperature (green) respectively. First order transitions occur in the regime $A_c<A< C$ and $T_c<T<T_{\mm{max}}\simeq 1.25T_c$. We have set $P\ell_0^5=5\,,C=10$.}
  \label{acoe}
\end{figure}

To proceed, we numerically compute the coexistence curve $A-T$. Notably, we reveal that the small-large black hole transition occurs for $T>T_c$, as shown in Fig. \ref{acoe}. This is substantially different from the charged AdS black hole case \cite{Cong:2021fnf,Cong:2021jgb}, in which the transition occurs for $T<T_c$, similar to the liquid-gas transition of Van-der Waals fluid. 

%In contrast, our holographic system is more akin to classical liquids than the quantum counterparts. 

Yet the transition temperature is upper bounded because of $A<C$, correponding to $\lambda>0$ ( the upper bound $\lambda=9/100$ corresponds to $A=C/2<A_c$, which is however located in the non-coexistence region). We evaluate the maxmum temperature analytically by performing Taylor expansions near this point
\be T_{\mm{max}}=\fft{3}{2^{7/20}(4+3\sqrt{2})^{3/10}}\,T_c\simeq 1.25 T_c \,.\ee

\subsection{C-charge criticality}
\begin{figure}
  \centering
  \includegraphics[width=210pt]{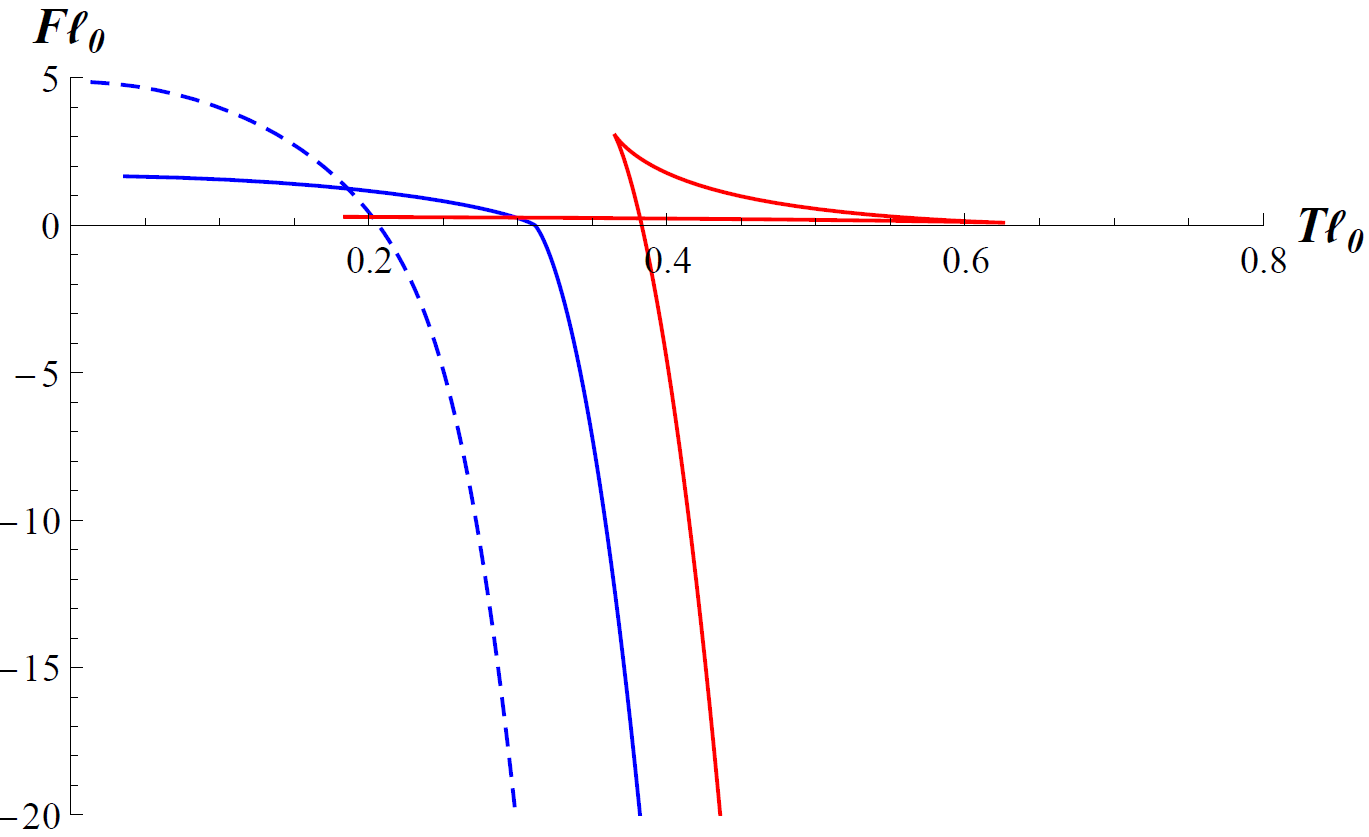}
    \includegraphics[width=210pt]{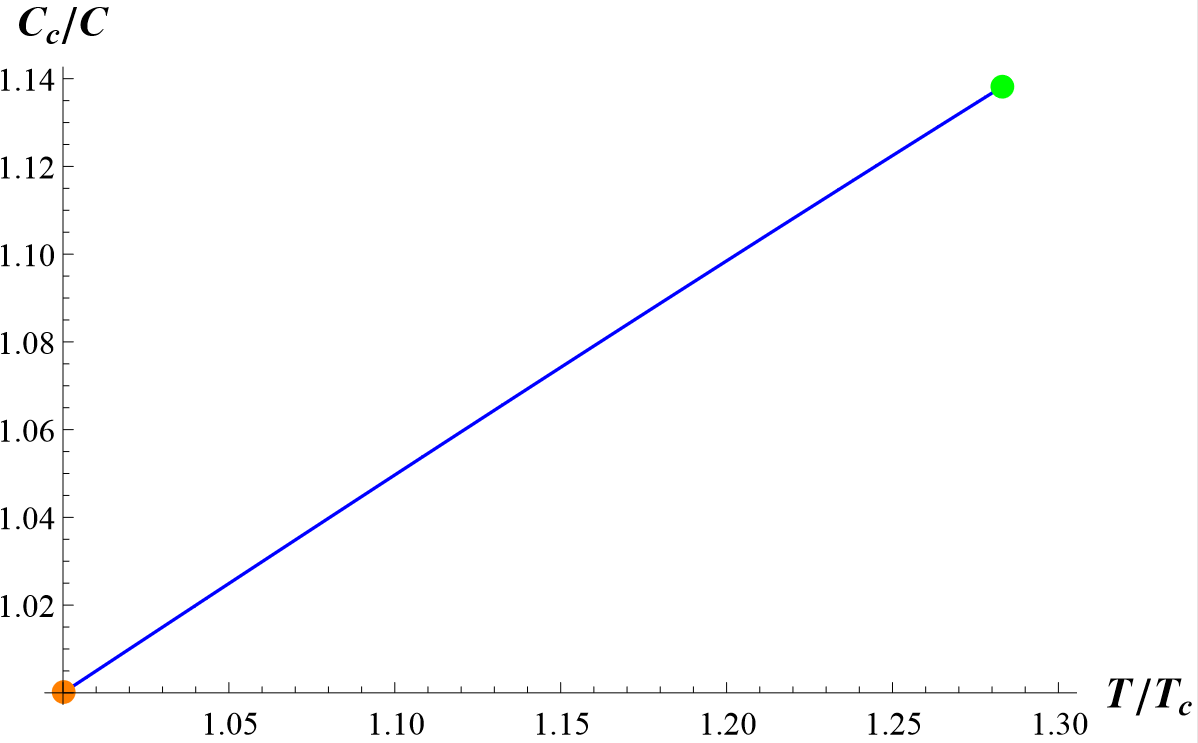}
  \caption{Left panel: $F-T$ diagram for $C/C_c=9/10$ (red), $C/C_c=1$ (blue) and $C/C_c=5/4$ (dashed). Swallow tail behavior emerges when $C<C_c$. Right panel: the coexistence curve. The first order transition occurs for $A<C<C_c$ and $T_c<T<T_{\mm{max}}\simeq 1.28 T_c$. We have set $P\ell_0^5=5\,,A=20$, where $\ell_0$ is an arbitrary reference length scale.}
  \label{ccrit}
\end{figure}
The $\mu_C-C$ criticality can be parallelly studied except for we work in the $(P\,,A)$ ensemble, corresponding to fixing volume and the $A$-charge in the boundary theories. We find
\bea
&& C_c=\fft{2+\sqrt{2}}{3}\,A\,,\nn\\
&& \ell_c=\fft{\sqrt{3(\sqrt{2}+1)}\,2^{3/10}  A^{1/5} }{ \pi^{2/5}(7+5\sqrt{2})^{3/10} P^{1/5} }\,.
\eea
To explore whether first order transion exists, we plot $F-T$ diagram for a given central charge $C$ at first, as shown in the left panel of Fig. \ref{ccrit}. Notably, swallower tail behavior emerges for central charge below the critical point $C<C_c$. Combining with previous results, we may conclude that the ratio between the numbers of degrees of freedoms characterized by the A-charge and the C-charge respectively plays a pivotal role for emergence of the Van der Waals-like behaviors in the boundary theories. This is similar to the charged AdS black holes case \cite{Cong:2021fnf,Cong:2021jgb}, in which critical phenomenon relies crucially on the ratio $C/Q^2$, where $Q$ is the electric charge in the boundary. This is in accordance with ordinary understanding of critical phenomenon in nature: the detailed balance between repulsive and attractive interactions in molecules essentially guarantees the existence of criticality as well as universality of critical exponents. 

Again, by numerically computing the coexistence curve, we establish that the first order transition occurs for $T>T_c$, as shown in the right panel of Fig. \ref{ccrit}. Again, the lower bound on the central charge $C>A$, corresponding to $\lambda>0$ implies a maximum transition temperature. We find
\be T_{\mm{max}}=\fft{3^{4/5}\sqrt{1+\sqrt{2}}}{2^{2/5}(5\sqrt{2}+7)^{3/10}}\,T_c\simeq 1.28 T_c \,.\ee
The result is close to but slightly different from the A-charge case.

\subsection{Critical exponents}
To proceed, we exmine critical exponents for central charges criticality. Without loss of generality, we focus on the $C$-charge (the $A$-charge case can be studied in the same way). By identifying $C\rightarrow P\,,\quad \mu\rightarrow V$, we immediately write down the definition for various exponents as 
\bea
&& C_\mu=T \fft{\pa S}{\pa T} \Big|_\mu\propto |t|^{-\alpha}\,,\\
&& \eta=\mu_g-\mu_\ell\propto |t|^\beta\,,\\
&& \kappa_T=-\fft{1}{\mu}\fft{\pa \mu}{\pa C}\Big|_{T}\,,\\
&& |C-C_c|\propto |\mu-\mu_c|^\delta\,,
\eea
where $t=T/T_c-1$. For later purpose, we introduce
\be \phi=\fft{\mu-\mu_c}{\mu_c}\,,\quad c=\fft{C}{C_c} \,.\ee
 Despite that we have a far more tedious equation of state $C=C(T\,,\mu)$, derivation of the critical exponents is straightforward. 

Firstly, since we work in $(P\,,A)$ ensemble and $C$-charge is fixed during the transition, the entropy $S$ will be a constant if $\mu$ is held fixed as well. This implies
\be C_\mu=0\quad\Rightarrow \quad \alpha=0  \,.\ee
Next, by expanding the equation of state near the critical point properly, one finds
\be\label{criteos} c=1-\fft{1}{2}\,t-\fft{3(2-\sqrt{2})}{4}\,t\phi+\fft{\sqrt{2}}{32}\,\phi^3+\mc{O}(t\phi^2\,,\phi^4) \,.\ee
Differentiating the equation yields 
\be dc=\Big( -\fft{3(2-\sqrt{2})}{4}\,t+\fft{3\sqrt{2}}{32}\phi^2\Big)d\phi\,.\label{dc}\ee
During the transition $c_g=c_\ell$ so that
\be\label{condition1} -\fft{3(2-\sqrt{2})}{4}\,t\phi_g+\fft{\sqrt{2}}{32}\phi^3_g=-\fft{3(2-\sqrt{2})}{4}\,t\phi_\ell+\fft{\sqrt{2}}{32}\phi^3_\ell \,.\ee
Besides, the Maxwell area law $\oint \mu dC=0$ implies
\be\label{condition2} 0=\int_{\phi_g}^{\phi_\ell}\phi\, dc=\int_{\phi_g}^{\phi_\ell}\phi\Big( -\fft{3(2-\sqrt{2})}{4}\,t+\fft{3\sqrt{2}}{32}\phi^2\Big)d\phi \,.\ee
The solution to (\ref{condition1}) and (\ref{condition2}) is 
\be \phi_g=-\phi_\ell=\sqrt{24(\sqrt{2}-1)t} \,.\ee
It should be emphasized that the solution makes sense only if $t>0$ (that is $T>T_c$), as desired. It is immediately seen that
\be \eta=\mu_c(\phi_g-\phi_\ell)\propto\sqrt{t}\quad\Rightarrow \quad \beta=1/2 \,.\ee
Furthermore, from (\ref{dc}), one finds
\be \kappa_T=-\fft{1}{C_c}\fft{\pa\phi}{\pa c}\Big|_{T}\propto -\fft{\sqrt{2}+1}{3t}\quad\Rightarrow \quad \gamma=1 \,.  \ee 
Finally, on the critical isotherm $T=T_c$, one has from (\ref{criteos})
\be c-1\propto \phi^3\quad\Rightarrow \quad \delta=3 \,.\ee
Unsurprisingly, the results are consistent with mean field theory and valid to the $A$-charge case as well.

\section{Conclusion }

In this work, employing extended phase space formalism, we explore critical phenomenon associated to the A-charge and the C-charge for holographic theories dual to Gauss-Bonnet black holes. In particular, we reveal an universal critical Gauss-Bonnet coupling, which leads to an universal ratio between the two central charges at the critical point. We may view the two central charges characerizing two distinct degrees of freedoms: one dominates attractive interactions and the other dominates repulsions. If this is correct, it implies that detaild balance in black hole molecules exists even if in the absence of electrically charged carriers in the boundary. It will be very interesing to study microscopic interactions for black hole molecules utilizing Ruppeiner geometry.

\section*{Acknowledgments}
The authors are supported in part by the National Natural Science Foundations of China with Grant No. 11805041 and No. 11873025 and also supported in part by Guangzhou Science and Technology Project 2023A03J0016.


\begin{thebibliography}{100}

%\cite{Bekenstein:1972tm}
\bibitem{Bekenstein:1972tm}
J.~D.~Bekenstein,
{\it Black holes and the second law,}
Lett. Nuovo Cim. \textbf{4}, 737-740 (1972).

%\cite{Bekenstein:1973ur}
\bibitem{Bekenstein:1973ur}
J.~D.~Bekenstein,
{\it Black holes and entropy,}
Phys. Rev. D \textbf{7}, 2333-2346 (1973).


%\cite{Bardeen:1973gs}
\bibitem{Bardeen:1973gs}
J.~M.~Bardeen, B.~Carter and S.~W.~Hawking,
{\it The Four laws of black hole mechanics,}
Commun. Math. Phys. \textbf{31}, 161-170 (1973).

%\cite{Maldacena:1997re}
\bibitem{Maldacena:1997re}
J.~M.~Maldacena,
{\it The Large N limit of superconformal field theories and supergravity,}
Adv. Theor. Math. Phys. \textbf{2}, 231-252 (1998)
doi:10.4310/ATMP.1998.v2.n2.a1
[arXiv:hep-th/9711200 [hep-th]].

%\cite{Witten:1998qj}
\bibitem{Witten:1998qj}
E.~Witten,
{\it Anti-de Sitter space and holography,}
Adv. Theor. Math. Phys. \textbf{2}, 253-291 (1998)
doi:10.4310/ATMP.1998.v2.n2.a2
[arXiv:hep-th/9802150 [hep-th]].


%\cite{Gubser:1998bc}
\bibitem{Gubser:1998bc}
S.~S.~Gubser, I.~R.~Klebanov and A.~M.~Polyakov,
{\it Gauge theory correlators from noncritical string theory,}
Phys. Lett. B \textbf{428}, 105-114 (1998)
doi:10.1016/S0370-2693(98)00377-3
[arXiv:hep-th/9802109 [hep-th]].


%\cite{Kubiznak:2012wp}
\bibitem{Kubiznak:2012wp}
D.~Kubiznak and R.~B.~Mann,
{\it P-V criticality of charged AdS black holes,}
JHEP \textbf{07}, 033 (2012)
[arXiv:1205.0559 [hep-th]].


%\cite{Gunasekaran:2012dq}
\bibitem{Gunasekaran:2012dq}
S.~Gunasekaran, R.~B.~Mann and D.~Kubiznak,
{\it Extended phase space thermodynamics for charged and rotating black holes and Born-Infeld vacuum polarization,}
JHEP \textbf{11}, 110 (2012)
doi:10.1007/JHEP11(2012)110
[arXiv:1208.6251 [hep-th]].


%\cite{Cheng:2016bpx}
\bibitem{Cheng:2016bpx}
P.~Cheng, S.~W.~Wei and Y.~X.~Liu,
{\it Critical phenomena in the extended phase space of Kerr-Newman-AdS black holes,}
Phys. Rev. D \textbf{94}, 024025 (2016)
doi:10.1103/PhysRevD.94.024025
[arXiv:1603.08694 [gr-qc]].


%\cite{Cai:2013qga}
\bibitem{Cai:2013qga}
R.~G.~Cai, L.~M.~Cao, L.~Li and R.~Q.~Yang,
{\it P-V criticality in the extended phase space of Gauss-Bonnet black holes in AdS space,}
JHEP \textbf{09}, 005 (2013)
doi:10.1007/JHEP09(2013)005
[arXiv:1306.6233 [gr-qc]].


%\cite{Cong:2021fnf}
\bibitem{Cong:2021fnf}
W.~Cong, D.~Kubiznak and R.~B.~Mann,
{\it Thermodynamics of AdS Black Holes: Critical Behavior of the Central Charge,}
Phys. Rev. Lett. \textbf{127}, no.9, 091301 (2021)
doi:10.1103/PhysRevLett.127.091301
[arXiv:2105.02223 [hep-th]].

%\cite{Cong:2021jgb}
\bibitem{Cong:2021jgb}
W.~Cong, D.~Kubiznak, R.~B.~Mann and M.~R.~Visser,
{\it Holographic CFT phase transitions and criticality for charged AdS black holes,}
JHEP \textbf{08}, 174 (2022)
doi:10.1007/JHEP08(2022)174
[arXiv:2112.14848 [hep-th]].

%\cite{Kumar:2022afq}
\bibitem{Kumar:2022afq}
N.~Kumar, S.~Sen and S.~Gangopadhyay,
{\it Breaking of the universal nature of the central charge criticality in AdS black holes in Gauss-Bonnet gravity,}
Phys. Rev. D \textbf{107}, no.4, 046005 (2023)
doi:10.1103/PhysRevD.107.046005
[arXiv:2211.00925 [gr-qc]].



%\cite{Brigante:2007nu}
\bibitem{Brigante:2007nu}
M.~Brigante, H.~Liu, R.~C.~Myers, S.~Shenker and S.~Yaida,
{\it Viscosity Bound Violation in Higher Derivative Gravity,}
Phys. Rev. D \textbf{77}, 126006 (2008)
doi:10.1103/PhysRevD.77.126006
[arXiv:0712.0805 [hep-th]].


%\cite{Camanho:2009vw}
\bibitem{Camanho:2009vw}
X.~O.~Camanho and J.~D.~Edelstein,
{\it Causality constraints in AdS/CFT from conformal collider physics and Gauss-Bonnet gravity,}
JHEP \textbf{04}, 007 (2010)
doi:10.1007/JHEP04(2010)007
[arXiv:0911.3160 [hep-th]].


%\cite{Nojiri:1999mh}
\bibitem{Nojiri:1999mh}
S.~Nojiri and S.~D.~Odintsov,
{\it On the conformal anomaly from higher derivative gravity in AdS / CFT correspondence,}
Int. J. Mod. Phys. A \textbf{15}, 413-428 (2000)
doi:10.1142/S0217751X00000197
[arXiv:hep-th/9903033 [hep-th]].


%\cite{Myers:2010jv}
\bibitem{Myers:2010jv}
R.~C.~Myers, M.~F.~Paulos and A.~Sinha,
{\it Holographic studies of quasi-topological gravity,}
JHEP \textbf{08}, 035 (2010)
doi:10.1007/JHEP08(2010)035
[arXiv:1004.2055 [hep-th]].

%\cite{Dutta:2022wbh}
\bibitem{Dutta:2022wbh}
S.~Dutta and G.~S.~Punia,
{\it String theory corrections to holographic black hole chemistry,}
Phys. Rev. D \textbf{106}, no.2, 026003 (2022)
doi:10.1103/PhysRevD.106.026003
[arXiv:2205.15593 [hep-th]].

%\cite{Punia:2023ilo}
\bibitem{Punia:2023ilo}
G.~S.~Punia,
{\it Bulk-Boundary Thermodynamics of Charged Black Holes in Higher Derivative Theory,}
[arXiv:2305.06552 [hep-th]].


%\cite{Myers:2010tj}
\bibitem{Myers:2010tj}
R.~C.~Myers and A.~Sinha,
{\it Holographic c-theorems in arbitrary dimensions,}
JHEP \textbf{01}, 125 (2011)
doi:10.1007/JHEP01(2011)125
[arXiv:1011.5819 [hep-th]].





\end{thebibliography}
\end{document}